\documentstyle[11pt]{article}
\def\ZZ{{\mathchoice {\hbox{$\sf\textstyle Z\kern-0.4em Z$}}
{\hbox{$\sf\textstyle Z\kern-0.4em Z$}}
{\hbox{$\sf\scriptstyle Z\kern-0.3em Z$}}
{\hbox{$\sf\scriptscriptstyle Z\kern-0.2em Z$}}}}

\newcommand{\ea}{\end{array}}

\begin{document}
\thispagestyle{empty}
\noindent\hspace*{\fill}  FAU-TP3-96/4 \\
\noindent\hspace*{\fill}  hep-th/9604099 \\
\noindent\hspace*{\fill}  April 17, 1996.  \\

\begin{center}\begin{Large}\begin{bf} 
Field Configurations and the SU(2) Haar Measure in QCD \\  
\end{bf}\end{Large}\vspace{.75cm}
 \vspace{0.5cm}
Harald W. Grie{\ss}hammer \footnote{hgrie@theorie3.physik.uni-erlangen.de}
 and Alex C. Kalloniatis \footnote{ack@theorie3.physik.uni-erlangen.de} \\
Institut f\"ur Theoretische Physik III \\
Universit\"at Erlangen - N\"urnberg \\
Staudtstra{\ss}e 7 \\
D-91058 Erlangen, Germany    
\end{center}
\vspace{1cm}\baselineskip=35pt
\begin{abstract} \noindent 
We characterise a class of SU(2) 
gluonic field configurations in the modified axial gauge where a zero mode
component vanishes at some space point but the global Haar measure remains
non-zero. The consequence of this is that gluonic wavefunctionals
need not vanish at the boundary of the fundamental modular domain, which
itself permits $\theta$ dependence in QCD(3+1). 
 
\end{abstract}
\newpage\baselineskip=18pt

It has been known for some time that quantum chromodynamics in
four dimensions must involve a hidden parameter known
variously as the $\theta$-parameter or the QCD vacuum angle \cite{JaR76}. 
The existence of this angle, analogous to the Bloch momentum
in a periodic potential, is related intimately to the presence
of large (topologically non-trivial) gauge symmetries in the theory.
In the Hamiltonian formalism, $\theta$ dependence creeps into
the theory due to the quasiperiodicity (periodic up to an
arbitrary phase) of the wavefunctional at the boundary
of the so-called Fundamental Modular Domain, the region of
gauge configuration space inside which unique representatives of gauge 
orbits have been identified. We shall further clarify this below.

The massive Schwinger model is the simplest field theory in
which $\theta$ dependence plays a role in the mesonic spectrum
\cite{CJS75}. However, QCD(1+1), the next step in non-triviality, 
has no $\theta$ dependence \cite{Wit79}. 
In the above terms, this can be explained as follows: the 
fundamental modular domain and a related region in gauge configuration
space, the Gribov region, happen to be identical. At the boundary
of the Gribov region, the Gribov {\it horizon},  
wavefunctionals turn out to {\it vanish}. Thus wavefunctionals
must vanish at the boundary of the fundamental modular domain, and there  
is no room left for $\theta$ to appear.

To proceed to higher dimensions now, we first recall  
why wavefunctionals happen to vanish at the Gribov horizon.
The space of gauge configurations in non-Abelian theories
is in general `curvilinear', which leads to a non-trivial 
Jacobian, a global operator, in the course of the elimination of 
redundant gauge field variables. 
This Jacobian can either sit in the kinetic term of the
gauge fields in the Hamiltonian, or be absorbed into the
wavefunctional -- much like the change to the radial basis
in the hydrogen atom (we are evidently working
in Schr\"odinger representation). The Gribov horizon is precisely
the sub-manifold in gauge space 
where the Jacobian first vanishes and hence,  
in the `radial' basis, the wavefunctional
vanishes on the Gribov horizon. Now in 3+1 dimensions,
one has a choice of gauges
to attempt to work in. The traditional Coulomb gauge is problematic
because it is impossible to perform the gauge fixing  
in closed form and has hindered explicit construction of the
fundamental modular domain. 
Nonetheless, one can show \cite{Coul} that the fundamental modular
domain is a proper subset of the Gribov region, though points
on the boundary of the domain may also lie on the Gribov horizon.
This shows that $\theta$ dependence is possible.  
In the static temporal gauge \cite{JLP91} and modified axial gauge 
\cite{Yab89,LNT94}, where the Jacobian is the gauge group
invariant Haar measure and {\it can} be written down 
although in an UV unregulated
state, one encounters a product over space points of, what we
shall call, a {\it local Jacobian}. Now a problem ensues:   
if the local Jacobian vanishes for any one space point, it would seem 
that the entire Jacobian must also vanish. If this also
happens to occur at the boundary of the fundamental modular domain,
which is not unlikely as will be clarified below, then
$\theta$ dependence is absent in QCD(3+1) in this gauge, a result in
contradiction to conventional wisdom as well as phenomenological
evidence \cite{Cre77}. In this letter, we show that
this sequence of logic need not hold.    
        
In order to make things more concrete we work in the Hamiltonian formulation
on the three-torus of periodicity $L$ 
in the modified axial gauge of \cite{LNT94}.
Our discussion can be easily translated to the static temporal gauge approach
of \cite{JLP91}. 
We moreover restrict ourselves to the case of SU(2) pure gauge theory,
though the generalisation of what follows is straightforward.
Thus the gauge is, in the first step, 
$\partial_3 A^a_3 = 0, a = 1,\dots,3$.
This is followed by an additional colour
rotation such that the surviving zero mode of $A_3$ is diagonal
in colour space.  

There are now two ways to express the Haar measure in functional space:
\begin{equation}
{\cal J}[a_3] = \prod_{x_\perp}^\infty \sin^2(\pi a_3(x_\perp))
\label{jacob1} 
\end{equation}
or 
\begin{equation}
{\cal J}[a_3] = \exp[\delta^{(2)}(0) 
\int d^2x_\perp \ln \sin^2( \pi a_3(x_\perp))] 
\label{jacob2} 
\end{equation}
where $a_3(x_\perp)$ is the dimensionless
combination $gL\sqrt{A^a_3(x_\perp) A^a_3(x_\perp)} /\pi$. It is thus dependent
only on time (suppressed in the Hamiltonian formalism) 
and the two-dimensional transverse
coordinates $x_\perp$, and is a scalar in colour SU(2) space.  
The first expression (\ref{jacob1}) allows us to 
clarify our earlier comments: we observe that the left hand side is a global
operator while the right hand side involves the `local
Jacobian' $\sin^2(\pi a_3(x_\perp))$, namely the 
Jacobian of gauge space at each space point separately. The argument of this
functional vanishes whenever dynamics lead the zero mode $a_3(x_\perp)$ itself
to be an integer. However, the wavefunctional $\Psi[a_3]$
should satisfy the boundary condition `$\Psi[0] = \exp(- i \theta) \Psi[1]$' 
(we suppress for simplicity the dependence on other variables   
which do not affect this line of argument).
As in the case of the radial variable in the hydrogen atom,
the standard kinetic energy term $-\delta^2/2\delta a_3^2$
is retrieved in the function space restricted by the condition that
$\Psi = 0$ when ${\cal J} = 0$. One can now verify the
following statements: (a) in the above gauge, the fundamental modular
domain is the restriction of the spatially constant part of $a_3$,
$\int d^2x_\perp a_3/L^2$, to the interval $[0,1]$, (b) residual
gauge transformations map the boundary point $0$ into the boundary point $1$,
and (c) for each configuration at the boundary of the fundamental modular
domain, $a_3 \in \ZZ$, the local Jacobian must vanish for
at least two space points $x_\perp$.      
We can now state the apparent problem precisely: 
if (i) $a_3 \in \ZZ$ at even one space point $x^{(0)}_\perp$, 
then because of the product structure in equation (\ref{jacob1})
it seems to follow that (ii) ${\cal J} = 0$ globally, 
thus (iii) $\Psi[0] = \Psi[1] = 0$ and no further role
can be played by $\theta$ in the theory. We shall now 
construct counterexamples to the suggestion that
(ii) should follow from (i) in higher dimensions.
               
Observe that equations (\ref{jacob1}) and
(\ref{jacob2}) are ill-defined expressions due to ultraviolet 
divergences. They actually exhibit a competition between ultralocality
(seen most evidently in equation (\ref{jacob1})) and non-locality. 
The question of choice of regulator is delicate. 
We shall avoid the lattice framework,
since in a finite lattice the entire subtlety is thrown into
the taking of the continuum limit. We also
avoid dimensional regularisation it usually being only 
perturbatively valid.  We are thus faced
with the actual problem: a choice of an alternative which
could respect symmetries, especially gauge symmetry.
For lack of such an alternative, let us simply argue what should
happen in an appropriate continuum
regularisation scheme. Firstly, the form of equations   
(\ref{jacob1}) and (\ref{jacob2}) is dictated by gauge symmetry,
so that in a scheme respecting this symmetry, these expressions
cannot change in structure. Secondly, in a nonperturbative
treatment the regulated form of 
the otherwise singular delta function in equation (\ref{jacob2})
should bring in some new scale which should be related to
$\Lambda_{\rm{QCD}}$. This is
essentially the argument of Johnson et al.\cite{JLP91}.
We can explain this
somewhat further by brief recourse to dimensional regularisation:
there, $\delta^{(2)}(0)$ is simply the tadpole integral in
momentum space.  In perturbation theory 
this can be set to zero because of the {\it absence}  
of any intrinsic scale in this regime. Nonperturbatively,
as mentioned, this is no longer the case and the integral
would be related to the appropriate scale.
We assume the same holds for any other nonperturbative
gauge invariant regularisation scheme.
We use for the moment the second expression (\ref{jacob2}),
and interpret the delta function as  
$\delta^{(2)} (0) = 1/\kappa^2$,   
with $\kappa$ some fundamental length scale in the theory
(similar to $a$ in \cite{JLP91}).
This scale would be important in a full dynamical calculation, but
in the following it plays no further role.   

Evidently $\Psi$ is nonzero as long as ${\cal J} \neq 0$, or,
using form (\ref{jacob2}), when  
\begin{equation} 
   \int d^2 x_\perp \ln \sin^2(\pi a_3(x_\perp))\;>\;-\infty\;\;.
\label{invest} 
\end{equation} 
Now, one notes that the Lesbegue integral in (\ref{invest}) is insensitive to
the exclusion of a countable number of points or lines on which the integrand
diverges. However, if $a_3(x_\perp) \in \ZZ$ for $x_\perp$
lying in a two-dimensional sub-{\it manifold} of 
the torus then the integrand over these points is infinite: for such  
configurations the Jacobian ${\cal J}$ really is zero, namely
the wavefunction vanishes. The configurations have zero probability
and play no role in the theory. In particular no $\theta$ dependence
can be associated with them. What is implicit in the above is that
we are thinking of wavefunctions which are essentially delta functionals
peaked about particular classical gauge field congurations,
$\Psi[a_3] \sim \delta[ a_3 - a_3^{\rm{class}}]$. Our task is to determine
examples of configurations $a_3^{\rm{class}}$ such that $\theta$
dependence can occur.     

We thus look for configurations $a_3(x^{(0)}_\perp) \in \ZZ$
for only one isolated point  $x^{(0)}_\perp$ on the transversal plane.
The neighbourhood of $x^{(0)}_\perp$ can be 
parametrised by polar coordinates $(r,\varphi)$, and (\ref{invest}) becomes
\begin{equation} 
   \lim\limits_{\varepsilon\to 0}\int\limits_{\varepsilon} dr\int d\varphi
              \;r \ln\sin^2(\pi a_3(r,\varphi)) \;>\;-\infty\;\;.
\label{limeps} 
\end{equation} 
For a finite integral, the integrand must behave like $r^{\alpha}$ with
 $\alpha>-1$,
\begin{equation} 
      r \ln\sin^2\pi a_3(r,\varphi) ={\cal A}(\varphi)\; r^{\alpha} +
      \mbox{ ... }\;\;,\;\;\alpha > -1\;\;\mbox{ as }r\to 0\;\; .
\end{equation} 
Solving for the field $a_3$ yields
\begin{equation} 
    \pi a_3(r, \varphi)=n\pi \pm\arcsin\exp\Big[\frac{{\cal A}(\varphi)}{2}\;
     r^{\alpha-1}\Big]\;\;\mbox{ as }r\to 0\;\;.
\label{twoperdim} 
\end{equation} 
Since $a_3(x^{(0)}_\perp)$  must be an integer,
\begin{equation} 
    - 1<\alpha<1\;\;\mbox{ and }\;\;{\cal A}(\varphi)<0\;\;.
\end{equation} 
The same calculation gives for a line singularity whose neighbourhood is now
parametrised by curvilinear coordinates, namely the
distance $r$ from the line and $s$, parametrising the line itself,
\begin{equation} 
    \pi a_3(r, s)=n\pi \pm\arcsin\exp\Big[\frac{{\cal A}(s)}{2}\;
r^{\alpha}\Big]\;\;,\;\;-1<\alpha<0\;,\,\;\;{\cal A}(s)<0\;\;\mbox{ as }r\to 0.
\label{oneperpdim} 
\end{equation} 
Noting that with the integrand in (\ref{invest}), also the integral cannot
become positive, the extension to several singularities is easily
accomplished by appropriately patching together solutions of
the types (\ref{twoperdim}) and (\ref{oneperpdim}). 
Since singularities cannot cancel, the integral about each
singularity has to obey (\ref{limeps}).
We note that the solution for point singularities approaches integer
values faster than (\ref{oneperpdim}), allowing for a larger
range of exponents $\alpha$. Neither (\ref{twoperdim}) nor (\ref{oneperpdim})
is analytic at $r=0$.
Finally, these expressions can be generalised to arbitrary dimensions,
for example if an appropriate nonperturbative generalisation of dimensional
regularisation can be validated. In $(d+1)+1$ dimensions, where $d$
is the dimensionality of the hypertorus on which $x_\perp$ lives,
we take a coordinate system with $r$ the distance of a point
from the hypersurface of singularities of dimension $N \leq d$, 
and with $(\varphi_1,\dots,\varphi_d)$ the
other coordinates. Then the required field configuration is
\begin{eqnarray}
\pi a_3(r, \varphi_1,\dots,\varphi_d )=n\pi \pm\arcsin\exp
\Big[\frac{{\cal A}(\varphi_1,\dots,\varphi_d)}{2}\;
 r^{\alpha-(d-N)}\Big]
\;\;, \nonumber \\  \;\;-1 < \alpha < d-N
\;;\,\;\;{\cal A}(\varphi_1,\dots,\varphi_d)<0\;\;\mbox{ as }r\to 0
\;.
\label{manyperpdims}
\end{eqnarray}
 
Now for each choice of $n$ for the configurations
(\ref{twoperdim}), (\ref{oneperpdim}) and (\ref{manyperpdims}), 
there exist under certain
conditions \cite{Gri96} configurations with $n$ replaced by $n+1$
corresponding to a transformation of winding number one.  
Moreover, each of these classical configurations lie precisely on the
boundary of the fundamental modular domain which itself corresponds to
the completely space-independent part of $a_3$ taking the value
zero or one. Thus, the quantum wavefunctionals corresponding
to these configurations must be identified up to an arbitrary
phase and this is where $\theta$ makes its appearance. 

Our construction relies on the existence of an infinite number
of degrees of freedom. 
Equation (\ref{jacob1}) shows that in any {\it quantum mechanical} 
framework `tunneling through the Jacobian' cannot occur.  
We stress moreover that (\ref{twoperdim}), 
(\ref{oneperpdim}) and (\ref{manyperpdims}) represent an
infinitely large {\it family} of gauge configurations parametrised
by the functions ${\cal A}$ and exponents $\alpha$. 
They are not a `set of measure zero'.           
Thus indeed, with these elements alone a rich class of
gauge configurations can be seen to exist. We do not claim
that this is even an exhaustive classification of such configurations.  
   
In order to proceed with dynamical calculations, the next step is 
to construct corresponding finite energy classical configurations of
transverse gluon fields. 
This is delicate due to structures in the Hamiltonian of 
\cite{LNT94} involving the mode $a_3$ in the denominator
and the transverse part of the Gau{\ss} operator in the numerator,
itself depending on the transverse gauge fields. Thus, finite
energy contributions can only come from those configurations
whose corresponding transverse gluon parts can balance
the centrifugal energy of the zero mode.
This work is still underway.
The main goal in this direction would be to use these
classical configurations finally as the basis of
a WKB calculation in order to solve the Hamiltonian.

\end{document}